\documentclass[conference]{IEEEtran}
\IEEEoverridecommandlockouts
\usepackage{cite}
\usepackage[frozencache, cachedir=minted-cache]{minted}
\usepackage{listings}
\usepackage{amsmath,amssymb,amsfonts}
\usepackage[noend]{algpseudocode}
\usepackage{algorithm}
\usepackage{graphicx}
\usepackage{forest}
\usepackage{textcomp}
\usepackage{hyperref}
\usepackage{xcolor}
\usepackage{tabularx}
\usepackage{enumitem}
\usepackage{verbatim}
\usepackage{url}
\usepackage{lipsum} 
\usepackage{caption}
\lstdefinelanguage{Solidity}{
    keywords={contract, function, returns, uint},
    keywordstyle=\color{blue}\bfseries,
    commentstyle=\color{gray},
    identifierstyle=\color{black},
    sensitive=false,
    morecomment=[l]{//},
    morecomment=[s]{/*}{*/},
    morestring=[b]",
    morestring=[b]',
}

\def\BibTeX{{\rm B\kern-.05em{\sc i\kern-.025em b}\kern-.08em
    T\kern-.1667em\lower.7ex\hbox{E}\kern-.125emX}}
\begin{document}
\captionsetup[table]{labelsep=colon, textformat=simple}

\newcommand{\linebreakand}{
  \end{@IEEEauthorhalign}
  \hfill\mbox{}\par
  \mbox{}\hfill\begin{@IEEEauthorhalign}
}

\title {Versioned Analysis of Software Quality Indicators and Self-admitted Technical Debt in Ethereum Smart Contracts with Ethstractor}

\author{
\IEEEauthorblockN{Khalid Hassan\IEEEauthorrefmark{1}, Saeed Moradi\IEEEauthorrefmark{1}, Shaiful Chowdhury, Sara Rouhani,}
\IEEEauthorblockA{Department of Computer Science, University of Manitoba, Winnipeg, Canada\\
Email: \{hassank2, moradis1, Shaiful.Chowdhury, sara.rouhani\}@umanitoba.ca}
\thanks{\IEEEauthorrefmark{1} These authors contributed equally to this work.}
}

\maketitle
\begin{abstract}
The rise of decentralized applications (dApps) has made smart contracts imperative components of blockchain technology. As many smart contracts process financial transactions, their security is paramount. Moreover, the immutability of blockchains makes vulnerabilities in smart contracts particularly challenging because it requires deploying a new version of the contract at a different address, incurring substantial fees paid in Ether. This paper proposes \textbf{Ethstractor}, the first smart contract collection tool for gathering a dataset of versioned smart contracts. The collected dataset is then used to evaluate the reliability of code metrics as indicators of vulnerabilities in smart contracts. Our findings indicate that code metrics are ineffective in signalling the presence of vulnerabilities. Furthermore, we investigate whether vulnerabilities in newer versions of smart contracts are mitigated and identify that the number of vulnerabilities remains consistent over time. Finally, we examine the removal of self-admitted technical debt in contracts and uncover that most of the introduced debt has never been subsequently removed.
\end{abstract}


\section{Introduction}
Software quality assurance efforts such as software maintenance, bug identification, and bug prediction have been extensively studied in the literature \cite{kononenko_investigating_2015}, \cite{palomba_toward_2019}. However, these efforts have mainly focused on traditional software rather than exploring behaviors in code deployed to the blockchain. This is primarily due to the decentralized nature of blockchain, as most studies require a history of changes to the codebase, which exists on centralized platforms used for traditional development like GitHub, but does not exist for blockchain networks.


Smart contracts are programs running on a blockchain network that applications interact with to read and update ledger data. They automate the execution of agreements or workflows and trigger subsequent actions when predetermined conditions are met. Smart contracts permit trusted transactions and agreements to be carried out among disparate, anonymous parties without the need for a central authority, legal system, or external enforcement mechanism. They render transactions traceable, transparent, and irreversible.
While smart contracts are running on trusted and secure blockchain networks, they are still pieces of code that can be analyzed for bugs and vulnerabilities \cite{rouhani2019security}. However, vulnerable smart contracts present new challenges compared to traditional software, such as:

\begin{enumerate}
    \item Financial loss: Vulnerable smart contracts can lead to financial loss if they contain vulnerabilities that allow malicious actors to exploit them. For example, a bug could enable an attacker to steal funds or manipulate the contract in unintended ways.

    \item Loss of trust: If a smart contract is found to be vulnerable, it can erode trust in the underlying blockchain platform and the applications built on top of it. Users may be reluctant to interact with smart contracts if they perceive them as unreliable or insecure.

    \item Immutability: Smart contracts are immutable and cannot be easily modified once deployed on the blockchain. Fixing bugs in a deployed smart contract may require deploying a new version, which can be complex and costly, especially if the contract manages significant assets or has many users.
\end{enumerate}

In this paper, we collect a dataset of versioned smart contracts and analyze the relationship between code metrics and vulnerabilities by examining their evolution across versions. We also study the removal of self-admitted technical debt, addressing a gap in previous research. Lastly, we investigate if vulnerabilities are resolved in newer versions of smart contracts. Specifically, this study addresses the following questions: How can different versions of a smart contract deployed on Ethereum be linked together (\textbf{RQ1})? Can code metrics serve as indicators for vulnerabilities in smart contracts (\textbf{RQ2})? Are vulnerabilities mitigated in the most recent versions of the smart contracts (\textbf{RQ3})? and Is self-admitted technical debt eventually resolved (\textbf{RQ4})?

This study makes the following contributions.
\begin{itemize}
\item We developed Ethstractor\footnote{https://github.com/tcdt-lab/ethstractor}, the first tool for extracting versioned smart contracts from the Ethereum blockchain, providing a valuable resource for analyzing smart contract development and enhancing security auditing.
\item We created a comprehensive dataset of versioned smart contracts, aiding our analysis and serving as a resource for future research in blockchain security and smart contract evolution.
\item We explored the efficacy of code metrics as indicators of vulnerabilities within smart contracts.
\item We assessed how versioning affects smart contract functionality and security, showing that vulnerabilities and technical debt patterns remain consistent over time.
\item We investigated developers' efforts to mitigate identified vulnerabilities across various contract versions.
\item We examined the management of self-admitted technical debt, specifically whether it is resolved over time or persists across contract versions.
\end{itemize}

\section{Related Works}

\subsection{\textbf{Vulnerability Detection}}
Numerous approaches have been introduced to tackle the problem of vulnerable smart contracts, including deep learning, genetic algorithms, machine learning, and static analysis. HajiHosseinKhani \textit{et al.} \cite{hajihosseinkhani_unveiling_2024} developed a genetic algorithm-based model, creating a benchmark dataset of Solidity source code samples and an analyzer that profiles smart contracts by identifying relevant data attributes and eliminating redundancies. Abdelaziz and Hobor \cite{abdelaziz_schooling_2023} implemented a semi-supervised learning application using graph neural networks (GNNs) to analyze Ethereum contract bytecode and identify vulnerable functions.

Soud \textit{et al.} \cite{soud_praioritize_nodate} proposed PrAIoritize, an automated tool for predicting smart contract bug priorities using bidirectional long short-term memory and feed-forward neural networks. Colin \textit{et al.} \cite{colin_integrated_2024} introduced a machine learning approach with a runtime opcode extraction algorithm for feature extraction and a trigram-based method for vectorization. Badruddoja \textit{et al.} \cite{badruddoja_making_2022} proposed a Naive Bayes algorithm to classify vulnerabilities in Ethereum smart contracts. Gao \textit{et al.} \cite{gao_smartembed_2019} created SMARTEMBED, a web service tool that helps Solidity developers find repetitive contract code and clone-related bugs using code embeddings and similarity-checking techniques.

Static analysis tools are another approach for vulnerability detection. He \textit{et al.}\cite{detectionOfVulnerabilities} classified detection methods into five categories under static analysis, as well as deep learning methods, concluding that no single method can cover all vulnerabilities. Ghaleb \textit{et al.}\cite{howEffectiveAreTools} found that static analysis tools like Mythril \cite{mythril} and the Remix IDE\footnote{https://remix.ethereum.org/} are widely used for vulnerability detection. However, their evaluation showed that popular tools often miss vulnerabilities they are expected to detect.
\subsection{\textbf{Code Metrics, and Technical Debt}}
The use of code metrics for maintenance and bug prediction is well-studied in software quality literature. Chowdhury \textit{et al.} \cite{shaifulBugs} determined that previous positive findings are unreliable due to improper use of k-cross validation. However, they identified positive correlations between metrics and bug proneness at different method sizes, suggesting the need for specialized models. Ebrahimi \textit{et al.} \cite{techDebt} investigated self-admitted technical debt in smart contracts and observed that, while not common across all contracts, it is prevalent in the most used ones, with a notable correlation between technical debt and code cloning.

\section{Collecting and Categorizing Versioned Smart Contracts}
The first step is obtaining a dataset of different versions of smart contracts deployed on the Ethereum blockchain. Available datasets do not categorize smart contracts by version; they simply group them in a single directory. A natural approach is to group contracts by name, but this is not possible due to the lack of essential metadata for the following reasons:

\begin{enumerate}
    \item Since anyone can deploy contracts to Ethereum with enough Ether, contracts with the same name but different deployers should be treated as distinct. This requires knowing the deployer's metadata (e.g. deployer's address).
    \item Even with the deployer's address, the deployment order of contracts cannot be determined without the deployment timestamp.
\end{enumerate}

Obtaining this metadata requires accessing the Ethereum ledger through a node, which demands significant computing power, storage, and time for synchronization (ranging from one week to several weeks). To avoid this delay, we used Etherscan\footnote{https://etherscan.io/}, which allows querying the Ethereum network via their API. This enabled us to bypass the synchronization step and perform our initial analysis without the resource commitment of setting up a node.

To collect different versions of smart contracts and link them together, a preliminary analysis of contract deployment schemes and behaviors is conducted. The transactions within the Ethereum blockchain were analyzed to narrow down the types of transactions requiring further examination to link different contract versions. This analysis identified three categories of transactions:

\begin{enumerate}
    \item Transactions for the transfer of Ether between one address and another.
    \item Transactions for deploying smart contracts.
    \item Transactions for interacting with smart contracts.
\end{enumerate}

Regular transactions for transferring Ether between two addresses typically occur between two parties as direct payments for various reasons. Since they do not involve smart contracts, they are of little relevance to our work. 

On the contrary, transactions for the deployment of smart contracts are crucial for identifying different versions of a contract. These transactions typically contain an empty \textit{To} field since the contract in question is yet to be created and has no address. Furthermore, the \textit{Input} field of the transaction contains the compiled bytecode of the transaction and the contract address is computed and registered in the \textit{contractAddress} field once the transaction is mined by a miner and included within a block. Thus, given the hash of a transaction, it is trivial to determine if it is a contract creation transaction as long as the aforementioned conditions are met.


While this enables the identification of contract creation transactions, another important aspect of a contract's deployment is its deployer. Since many unrelated contracts may have the same name across the Ethereum network, we identify a unique contract by its name, contract address, and deployer's address combined. While analyzing different contract creation transactions, two deployment schemes were identified
. The first scheme involves a user constructing the transaction with their address as the origin and broadcasting it to the network. For contracts deployed using this scheme, their versions can be identified by first identifying the deployer as the origin address of the creation transaction. Then, by iterating over the deployer's transaction list, one can find other creation transactions for contracts with the same name.

The second deployment scheme involves a smart contract that acts as an intermediary for the deployment of the target smart contract. This scheme, known as \textit{Nick's Method} or \textit{Keyless Execution}, was first proposed by Nick Johnson \cite{nickmethod}. This scheme leverages the fact that when choosing a random Ethereum address, it is extremely unlikely that this address is active and is being used by some participants in the network. So, instead of computing signatures for transactions (which takes a long time for large amounts of transactions), the scheme involves filling the signature field of the transaction with a random yet predictable value such that the address that corresponds to that signature will likely not be controlled by another party. This is done through \textit{CREATE} contracts, usually to ensure that a contract can be deployed to multiple Ethereum-based blockchains with the same address in each network. 

Identifying the deployer in this case is tricky. For some contracts, the Etherscan API is intelligent enough to link the target contract to the original deployer's address even though it was deployed by a \textit{CREATE} contract. However, for contracts that are not linked by Etherscan, we first check if the \textit{From} field in the transaction contains the address of a contract. If the address was present, we then find the transaction responsible for initiating the creation of the target contract. This is done by iterating over the incoming transactions and matching the compiled bytecode included in the \textit{Input} field with the compiled bytecode of the deployed contract. If a match is found, the first four bytes of the \textit{Input} field-- which contain the \textit{Keccak-256} hash of the called method's signature-- are extracted. The intermediary contract's Application Binary Interface (ABI) is then used to match the hash to the name of the method that was called from it. If the name of the method contains the words \textit{deploy} or \textit{create}, and the contract's bytecode is included in the Input field of the transaction, this indicates that the transaction was used to create the target contract.

The final category of transactions is one that is used for interacting with smart contracts. From this category, in addition to transactions involving \textit{CREATE contracts} previously discussed, transactions involving \textit{proxy contracts} are essential to our work. Proxy contracts are similar to \textit{CREATE} contracts in that they are intermediary contracts for another contract. However, proxy contracts act as contracts with permanent end-user addresses, off-loading the execution logic to different contracts. This allows for deploying different versions of a contract without changing the address that end-users have to interact with. Proxy contracts serve as an information hub for our work since they interact with all different contract versions. Etherscan identifies such contracts, and we use that to iterate through their transactions and find the different contract versions.

\begin{figure}[htb]
    \centering
    \includegraphics[height= 120px, width=250px]{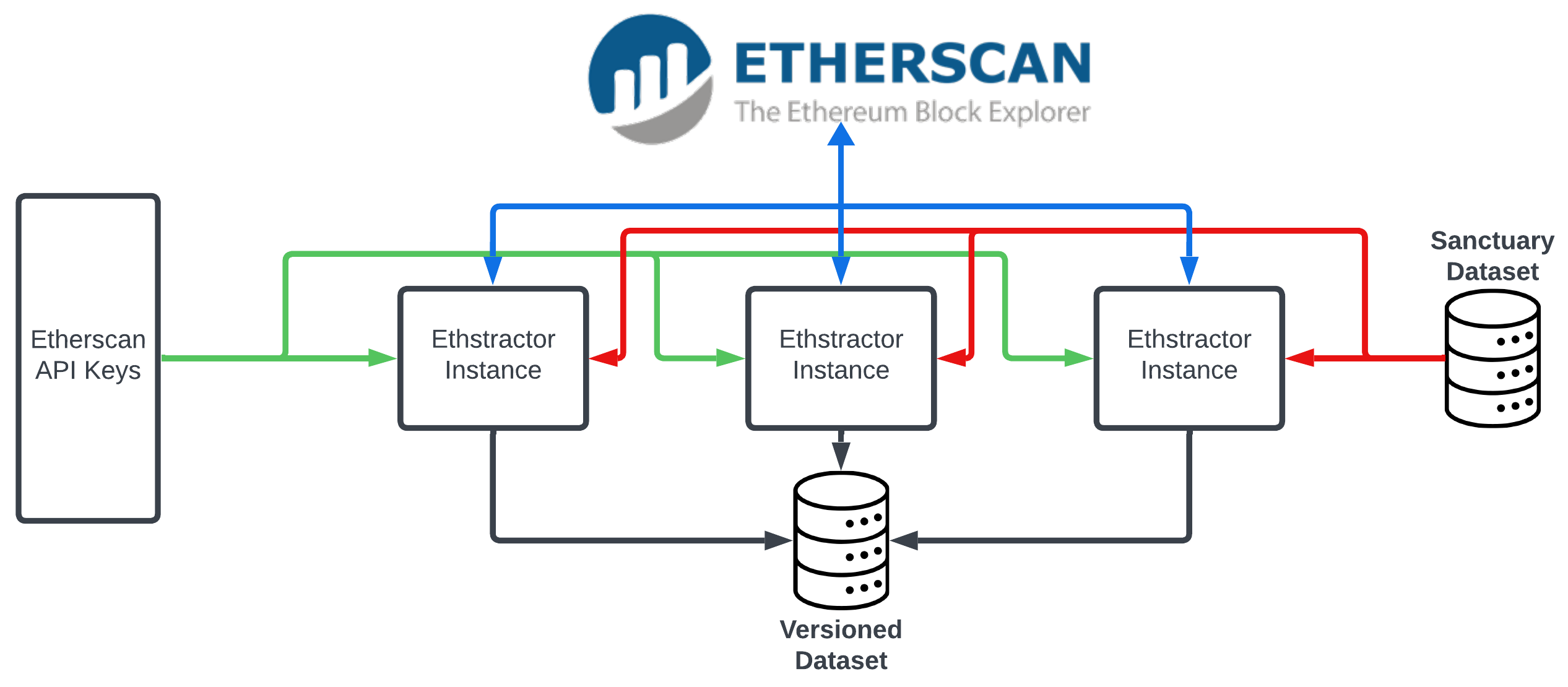}
    \caption{Architecture diagram showing the setup of the Ethstractor deployment. Blue lines signify interactions between Ethstractor and Etherscan; red lines are used for interactions between the Sanctuary dataset and Ethstractor; green lines show the reading of Etherscan API keys; and black lines show the writing to the output versioned dataset.}
    \label{fig:architecture}
\end{figure}

Using the previous analysis results, we propose \textit{\textbf{Ethstractor}}, a tool for extracting versioned smart contracts from the Ethereum blockchain. We chose not to build the tool in the form of a crawler that crawls through the blockchain and extracts different smart contract versions as it comes across them. This was done for two reasons: 
\begin{enumerate}
    \item Since we depend on Etherscan for the querying of the Ethereum blockchain, we are limited by their imposed rate limits for each API key. Crawling through the blockchain will require the usage of a large number of requests and will not result in finding contracts for the most part since most transactions are between addresses, resulting in our assigned rate limit being depleted without significant results.
    \item Finding contracts by crawling through the blockchain will require a significant amount of time to collect a dataset of sufficient size, which is not feasible given the time constraints of this work.
\end{enumerate}
To satisfy the time requirements of our work, we chose to use a dataset as a reference for our tool. Ethstractor has the non-versioned dataset as an input and iterate through it, extracting the address of each contract it iterates over and finding all of its different versions on Ethereum through Etherscan using the previously discussed strategies. We chose the Sanctuary dataset \cite{smart_contract_sanctuary}, which contains smart contracts collected from different Ethereum-based networks. Specifically, we used the \textit{mainnet} dataset since it only contains contracts deployed to the main Ethereum network. Furthermore, all contracts extracted within the Sanctuary dataset are named according to the following format:

\begin{minted}{bash}
contract_address_contract_name.sol
\end{minted}
This facilitates the identification of the contract on the Ethereum network, which in turn provides access to all associated metadata, including its transaction list, creation transaction, and deployer’s address.

Figure \ref{fig:architecture} shows the architecture of Ethstractor, implemented in Golang. Initially, the Sanctuary dataset is cloned to the remote server where Ethstractor is deployed. Ethstractor then navigates through the dataset's directories, extracting each contract’s address from its filename according to the format used by Sanctuary. Subsequently, it executes queries on Etherscan using the extracted addresses to collect all associated versions of each contract. While Etherscan accepts requests without API keys, it limits such requests to one every five seconds. However, including an API key in the URL-encoded request parameters increases the rate limit to five requests per second. Despite still being slow, this rate limit is adequate for our work, but another problem arises as each API key only allows 100,000 daily requests. With each query about a contract necessitating a request being sent to Etherscan, and with each contract requiring tens or hundreds of queries, this limit is insufficient.

To address this, we first optimized our queries to ensure only necessary queries are sent, and each query maximizes the amount of information retrieved (e.g., retrieving 50,000 transactions per query instead of only 10,000, as 50,000 is the response length limit set by Etherscan). While these optimizations greatly improved the utilization of API key rate limits, the tool still inevitably exceeded them. Therefore, to ensure the tool never has downtime due to rate limits, we implemented a mechanism to cycle through keys once the key currently being used exceeds its rate limit. Additionally, each account is allowed three different API keys, with rate limits applied on a per-key basis. Using this information, 18 different API keys were generated and stored in a file accessible to the tool, allowing it to cycle through them whenever it hits a rate limit.

While rate limits no longer posed a challenge, the rate at which the tool could extract different versions of unique contracts became the limiting factor. Running a single instance of Ethstractor yielded about 2,200 unique contracts per day, which is too slow to collect a sufficiently large dataset. This led us to deploy two additional instances on the remote server, with each of the three instances responsible for part of the Sanctuary dataset after it was equally divided into three parts. Since each contract has a unique address, even if a contract shares the same name, it is treated as a different file, thus eliminating any concerns about race conditions for writing the contracts to the output dataset. The three instances running in parallel were able to collect 159,105 unique contracts, each with one or more versions, over 18 days, equating to a rate of about 8,839 contracts per day.

Ethstractor separates each contract written into the dataset by name first and then further separates them by deployer since multiple deployers may have contracts that have the same name. The contract is then appended to the dataset with a file name that contains its address, name, and version according to the format shown in the figure. Finally, to increase the accessibility and the availability of the dataset, we created a script that automatically pushes all changes to a GitHub repository, where this script is periodically run (at the top of every hour) using a \textit{cron} job that was set up on the remote server on which Ethstractor was deployed.


Further analysis of the dataset revealed that some contracts have an anomalously high number of versions. For example, while most contracts have single-digit versions, the contract \textit{LockToken} has over 23,000 versions. Contracts with similar names serve a similar functionality of locking tokens for exchange and have many versions. These versions differ only in a single line in their constructor, which changes according to the nature of the transaction taking place, and a new version is deployed for each of these transactions. Despite understanding the reason behind this high number of versions, these contracts are of no use to this work since they are usually very simple and their versions are almost identical. Therefore, we chose to exclude all smart contracts with more than 100 versions, removing 64 contracts from the dataset. Our analysis demonstrates that most smart contracts have a small number of versions, with over 80\% of the contracts having only one version.

\vspace{10pt}
\fbox{\begin{minipage}{0.9\linewidth}
\textbf{Answer for RQ1:} Smart contract versions can be linked together by finding their deployer and iterating through their transactions to find the ones responsible for the creation of different versions of contracts.
\end{minipage}}


\section{Evaluating the Code Metrics in Smart Contracts}
After extracting the smart contracts, we built a tool to extract their code metrics. Below is the list of all the code metrics considered:

\begin{itemize}
    \item \textbf{SLOC (Source Lines of Code)}: The number of lines a piece of code consists of, excluding whitespaces and comments.
    \item \textbf{McCabe (Cyclomatic Complexity)}: The number of all decision points within a code snippet plus one. Examples of decision points are \textbf{if, while}, and \textbf{for}.
    \item \textbf{Halstead Volume}: A code metric that assesses program size by considering the count of distinct operations and variables utilized.
    \item \textbf{Maintainability Index}: A metric measuring the clarity and ease of maintenance of a program. Higher values on the maintainability index are preferred. The formula for this metric is provided below: \\
    \begin{equation}
        171 - 5.2 \cdot ln(HV) - 0.23 \cdot CC - 16.2 \cdot ln(SLOC) \label{maintainabilit_formula}
    \end{equation}

Throughout this paper, we abide by the following terminology when referring to these metrics:

\begin{table}[h]
\centering
\begin{tabular}{|c|c|}
\hline
\textbf{technical term} & \textbf{acronym used in the paper} \\
\hline
Source Lines of Code & SLOC \\
\hline
Cyclomatic Complexity & McCabe \\
\hline
Halstead Volume & HV \\
\hline
Maintainability Index & MI \\
\hline
\end{tabular}
\caption{Terminology used throughout the paper}
\label{tab:example_two_columns}
\end{table}

Code metric analysis can be conducted at two levels: method-level and file-level, both of which are considered in this study. To facilitate this analysis, a dedicated tool has been developed to compute these metrics. The reliability and accuracy of the tool's results are ensured through manual verification. Prior to discussing the experimental findings generated by the tool, it is imperative to elucidate certain pivotal aspects of its implementation.

\end{itemize}
\subsubsection{\textbf{Code Metrics Correlation Evaluation}}
After evaluating the code metrics in our smart contracts, we calculate their correlation coefficient to see if they are highly correlated with each other.

\subsubsection{\textbf{Vulnerability Evaluation}}
To attest the vulnerability of smart contracts, we used Slither\cite{feist_slither_2019},
a popular static analysis framework for Solidity smart contracts designed to identify security vulnerabilities, design issues, and potential bugs.

\subsubsection{\textbf{Vulnerability Correlation Evaluation}}
After detecting the vulnerabilities in smart contracts, the correlation coefficient between the total number of vulnerabilities present in a contract and each code metric is calculated.

\subsection{\textbf{Code Metrics Correlation Evaluation}}
We calculated the correlation coefficients between each pair of code metrics using \textit{Pearson, Spearman} and \textit{KendallTau} algorithms.

\subsubsection{\textbf{File-level}}
The results for the file-level evaluation are shown in table \ref{file-level-correlation-coefficients}.
\begin{table}[htbp]
    \centering
    \begin{tabular}{|c|c|c|c|}
        \hline
        \textbf{Code Metrics} & \textbf{Pearson} & \textbf{Spearman} & \textbf{Kendall Tau} \\
        \hline
        \textbf{SLOC-McCabe} & 0.81752 & 0.88996 & 0.76131 \\
        \hline
        \textbf{SLOC-HV} & 0.94434 & 0.98342 & 0.92833 \\
        \hline
        \textbf{SLOC-MI} & -0.72213 & -0.99724 &-0.98171 \\
        \hline
        \textbf{McCabe-HV} & 0.8907 & 0.90204 & 0.77568 \\
        \hline
        \textbf{McCabe-MI} & -0.70711 & -0.89739 & -0.77012 \\
        \hline
        \textbf{HV-MI} & -0.74195 & -0.99003 & -0.94658 \\
        \hline
    \end{tabular}
    \caption{file-level correlation coefficient between code metrics}
    \label{file-level-correlation-coefficients}
\end{table}

Based on the definition, anything above 0.8 is considered a strong correlation. According to the table \ref{file-level-correlation-coefficients}, every pair of code metrics demonstrates an absolute correlation of 0.7, denoting a strong correlation (either negative of positive) between one another. The MI has a high but negative correlation with other metrics since a maintainable code is a code with fewer lines (SLOC), fewer decision points (McCabe), and less complexity (HV). On the other hand, McCabe has the weakest correlation with other metrics.


\textbf{Why does McCabe have a weaker correlation compared to other metrics?}\\
If we look at the skeleton of Solidity smart contracts, we can see that there are numerous methods with few, if not zero lines of code that are written either within solidity \textit{interfaces}, \textit{modifiers}, or are simply getters\footnote{Methods used to retrieve the values of private or protected attributes (properties) of an object in object-oriented programming.} and setters \footnote{Methods used to modify the values of private or protected attributes (properties) of an object in object-oriented programming.} of Solidity classes. This means that a lot of these functions have no decision points to count when calculating McCabe, therefore their McCabe is equal to 1. 
The results are not surprising because in \cite{giger_method-level_2012} and \cite{di_nucci_developer_2018}, they discovered that code metrics in languages like Java are highly correlated to each other. Our results show that this holds true for a language like Solidity and smart contracts in a nutshell.



\subsubsection{\textbf{Method-level}}
The results for method-level evaluation are shown in table \ref{method-level-correlation-coefficients}.

\begin{table}[htbp]
    \centering
    \begin{tabular}{|c|c|c|c|}
        \hline
        \textbf{Code Metrics} & \textbf{Pearson} & \textbf{Spearman} & \textbf{Kendall Tau} \\
        \hline
        \textbf{SLOC-McCabe} & 0.23052 & 0.50030 & 0.43257 \\
        \hline
        \textbf{SLOC-HV} & 0.72747 & 0.72967 & 0.58655 \\
        \hline
        \textbf{SLOC-MI} & -0.31091 & -0.93310 & -0.82803 \\
        \hline
        \textbf{McCabe-HV} &0.60925 & 0.59700 & 0.50896 \\
        \hline
        \textbf{McCabe-MI} & -0.48836 & -0.56469 & -0.47034 \\
        \hline
        \textbf{HV-MI} & -0.54172 & -0.91639 & -0.77827 \\
        \hline
    \end{tabular}
    \caption{Method-level correlation coefficient between code metrics}
    \label{method-level-correlation-coefficients}
\end{table}

Unlike file-level evaluation, we observe a poor correlation in almost every pair of metrics except for SLOC-MI. This raises several questions, which we discuss below.

\textbf{Why is the same correlation not observed in the method-level evaluation?}\\
Although this might seem counter-intuitive at first, the explanation is similar to why we observe weaker correlations between McCabe and other metrics at the file-level analysis. Most methods are relatively simple and lack complexity, making it challenging to extract any clear patterns between the metrics of a given method. Typically, only a few methods in a single smart contract contain statements that perform significant functionalities. Consequently, method-level evaluation in smart contracts is not reliable due to the nature of the language in which they are implemented.
To further validate this claim, we analyzed the SLOC for all the methods we extracted. We found that 83\% of the methods contain only one line of code, which likely results in their McCabe score being equal to one (no decision points).

\textbf{Why is there a strong correlation between SLOC and MI?}\\
The MI \ref{maintainabilit_formula} clearly depends on the other three code metrics discussed in this study. Since other code metrics are quantitatively negligible, SLOC has the highest impact on the MI. As a result, these two metrics are highly correlated with one another.

\subsection{\textbf{Code Metrics and Vulnerability}}
For this section of our study, about 5800 smart contracts were considered for vulnerability detection.
Slither was run on each contract to detect the total number of vulnerabilities present, which was then used to calculate the correlation coefficient for each code metric. Mathematically, we followed the formula below for vulnerability calculation:

\begin{equation}
    V_{sc} = \sum_{i=1}^{N} t_{i}
\end{equation}

Where $V_{SC}$ refers to the vulnerability of a smart contract, $N$ is the total number of vulnerabilities in a smart contract, and $t_i$ is the amount of threat a vulnerability poses on a smart contract. For this study, we assume that this amount is equal to 1 for all the vulnerabilities detected in the smart contract. Therefore, we arrive at the following formula:

\begin{equation}
    V_{sc} = N_{SC}
\end{equation}

The table below demonstrates our final results from a file-level perspective. As discussed earlier, the method level of this study is not be reliable, so it is not considered throughout the rest of this work.

\begin{table}[htbp]
    \centering
    \begin{tabular}{|c|c|c|c|}
        \hline
        \textbf{Code Metrics} & \textbf{Pearson} & \textbf{Spearman} & \textbf{Kendall Tau} \\
        \hline
        \textbf{SLOC-Vulnerability} & 0.45992 & 0.40988 & 0.35441 \\
        \hline
        \textbf{HV-Vulnerability} & 0.46956 & 0.33236 & 0.31039 \\
        \hline
        \textbf{McCabe-Vulnerability} & 0.43158 & 0.38001 & 0.32330 \\
        \hline
        \textbf{MI-Vulnerability} & -0.45085 & -0.39957 & -0.35328 \\
        \hline
    \end{tabular}
    \caption{file-level correlation coefficient between code metrics and smart contract vulnerability}
    \label{tab:my_table}
\end{table}

The unexpected result indicates that none of the code metrics exhibit a strong correlation with the vulnerability of a smart contract to attacks. If we look at the correlation coefficients between MI and vulnerability, which represents the weighted correlation between all of our code metrics (according to \ref{maintainabilit_formula}), the values for the three algorithms are -0.4 on average. This suggests that code metrics cannot be good indicators of vulnerabilities in a smart contract.

\vspace{10pt}
\fbox{\begin{minipage}{0.9\linewidth}
\textbf{Answer for RQ2:} Code metrics are not good indicators of vulnerabilities in smart contracts since there is no strong correlation between them. Therefore, code metrics are not effective in predicting vulnerable smart contracts prior to release.
\end{minipage}}
\vspace{10pt}

Considering the obtained results, the absence of correlation requires further investigation. To explore this, we conducted an analysis of various versions of contracts to ascertain any observable patterns in their vulnerabilities over time. Specifically, we aimed to determine whether the vulnerabilities change in different versions of contracts. For that, we selected the contracts with more than 50 versions to ensure that we could observe their changes over a long period of time. This also ensured that the contracts had time to address their vulnerabilities. We found that the smart contract vulnerabilities do not change at all over time.  In our dataset, there was a smart contract named MultiSigStub\footnote{Deployer Address: \\
0x89bb5a1880608fe606f6d2c3dc30c3624f3429fc\\
Contract Address:\\ 0x004de0313fd383c166b6f4390f1ba6c476c505d1\_MultiSigStub\_V640\\} with 647 different versions, but they all had the same amount of vulnerability.\\
This observation suggests that versioning in smart contracts is not related to their vulnerabilities but rather to something else, like adding features or updating the smart contract to be compatible with the most recent blockchain networks.

For demonstration, the amount of vulnerabilities present in the smart contracts with more than 50 versions is shown in figure \ref{smart_contract_vulnerability_histogram}. 

\begin{figure}[htb]
    \centering
    \includegraphics[width=0.8\columnwidth]{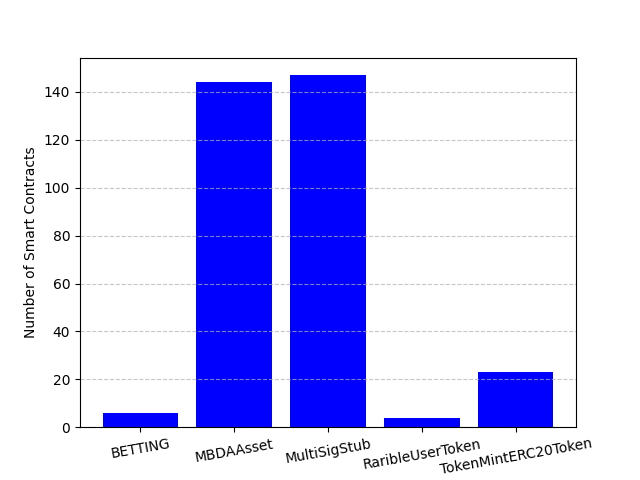}
    \caption{versioning in smart contracts}
    \label{smart_contract_vulnerability_histogram}
\end{figure}

\vspace{10pt}
\fbox{\begin{minipage}{0.9\linewidth}
\textbf{Answer for RQ3:} Smart contract vulnerabilities remain unchanged over time, which suggests that developers are either unaware of the vulnerabilities, unable to mitigate them, or ignore them.
\end{minipage}}

\section{Self-admitted Technical Debt in Smart Contracts}

\begin{table*}[htbp]
\centering
\begin{tabular}{|l|l|l|}
\hline
\textbf{Deployer Address} & \textbf{Contract Name} & \textbf{Sample Comment} \\ \hline
0xf5323d & A2Claim & caution: this function is deprecated because it requires allocating memory for the error \\ \hline
0x0de519 & ACL & // todo: this should be external \\ \hline
0x0de519 & ACL & // todo: add execution times param type? \\ \hline
0x2fb7d6 & AMMWithdrawer & // fix gas stipend for non standard erc20 transfer in case token contract's safemath violation is triggered \\ \hline
0xbf3d6f & AAVEInterestERC20 & // workaround to omit usage of abicoder v2 \\ \hline
\end{tabular}
\vspace{2mm}
\caption{Sample comments taken from different smart contracts.}
\label{tab:comment_table}
\end{table*}

Since previous work did not have access to a versioned dataset of smart contracts deployed on Ethereum, no studies explored the removal of self-admitted technical debt (referred to as technical debt from here on out for simplicity) over time, so we decided to fill that gap in the literature. To study technical debt removal, it must first be identified in the source code of the smart contracts present in the dataset. To do this, we first curated a list of keywords to identify comments that point towards the presence of technical debt. This was done by collecting keywords that have been used in the previous works \cite{techDebt}, \cite{potdar_exploratory_2014}. Furthermore, a brief manual analysis of smart contracts was carried out to identify more keywords that can potentially be added to the list, which contains the words \textit{todo, fix, fixme, deprecated, refactor, temporary, wip, work in progress, and workaround}.

Using this word list, comments within the source code of contracts which contain any of the keywords were extracted. For each contract, the evolution of its technical debt throughout its versions was recorded, as well as the comments signifying the presence of the technical debt of which a sample is shown in Table \ref{tab:comment_table}. Furthermore, the evolution of technical debt was measured by counting the instances of comments that point to technical debt in the initial version and then tracking their removal across subsequent versions or the addition of new technical debt. However, as mentioned in \cite{potdar_exploratory_2014}, it is important to check for inconsistent changes, such as removing a comment without changing the source code or the source code without removing the associated comment. 

This was achieved by extracting code snippets linked to each comment and assessing whether modifications were made to the snippets when their associated comments were removed. We also examined if changes occurred while the related comments remained unchanged. However, attributing any such change directly to the resolution of technical debt is not definitive. Therefore, in these instances, we opted to rely on the developers' comments, assuming that technical debt persisted despite any code modifications. Moreover, cases where comments were removed without corresponding changes in the linked code were not considered as technical debt resolutions. Our dataset did not contain any such instances, eliminating the need for further processing.

\begin{table}[htbp]
\centering
\begin{tabular}{|l|l|}
\hline
\textbf{Metric} & \textbf{Value} \\ \hline
Mean Initial Debt & 0.424 \\ \hline
Median Initial Debt & 1 \\ \hline
\% of Contracts with Debt Removal & 6.49\% \\ \hline
\end{tabular}
\vspace{2mm}
\caption{Self-admitted technical debt statistics.} 
\label{tab:debt_table}
\end{table}

Table \ref{tab:debt_table} shows statistics regarding the self-admitted technical debt found in the dataset. The mean technical debt, being 0.424, shows that most contracts do not contain self-admitted technical debt, and if they do, it will be a single instance of debt as indicated by the median. This corroborates the findings in \cite{techDebt} in that most contracts do not contain instances of self-admitted technical debt. Moreover, we found that technical debt is removed in only 6.49\% of the contracts which had technical debt introduced to it in its initial version. This means that while technical debt is removed in some contracts, the technical debt introduced to most contracts remains the same. 

\vspace{10pt}
\fbox{\begin{minipage}{0.9\linewidth}
\textbf{Answer for RQ4:} Most self-admitted technical debt introduced to smart contracts is never removed after its introduction, with some contracts even increasing the amount of self-admitted technical debt in their subsequent versions.
\end{minipage}}

\section{Threats to validity}

The \textbf{Internal Validity} of this study can be criticized for the correctness of the source code retrieved for contracts from Etherscan. For source code to be available on Etherscan, the contract's deployer has to provide a version of the source code, which then gets compiled by Etherscan to compare the resulting bytecode with the bytecode present in the blockchain. Using these verified source codes, Etherscan uses a similarity search and matches nearly identical bytecode instances to verified source codes. While this means that the obtained source for some contracts may not be fully accurate, the difference is usually only in a single line in the constructor (contracts with a large number of versions which have been eliminated from our dataset) of the contract, which does not affect any of our results regarding code metrics or the evaluation of technical debt.

Furthermore, the identification of comments indicating self-admitted technical debt is also a concern since our work trusts the developers' comments, and we curate a list of keywords for the identification of the comments. However, the keywords used were picked according to ones that were successfully used in literature and according to our brief manual analysis of the smart contracts.

The External Validity of this study can be criticized in terms of the generalizability of its results, given that our dataset contains 159,105 unique contracts, which is considerably smaller than other publicly available (non-versioned) datasets. This limitation arises from the tight time frame in which the dataset was collected. However, this can be easily mitigated by extending the data collection period. Additionally, smaller subsets of the dataset were used for code metric evaluation and smart contract vulnerability detection, comprising 22,279 and 5,762 contracts, respectively. This reduction was primarily due to time constraints, as even processing this smaller number of smart contracts required over 15 hours due to the necessity of installing different Solidity versions for each contract before analysis with Slither. Nonetheless, we believe these subsets are representative of the entire dataset, as they encompass a diverse range of contracts with varying numbers of versions.



To quantitatively measure vulnerabilities in smart contracts, we calculated the total number of vulnerabilities to compare with code metrics. Despite the diversity of vulnerabilities, we aggregated them under the assumption that each type equally contributes to the overall threat posed by a smart contract. This simplification warrants further investigation, as a more nuanced analysis should categorize vulnerabilities by type and impact. If vulnerabilities are indeed evenly distributed across the identified types, treating them with equal importance would be justified. However, a detailed categorization could reveal differences in severity and potential impact, necessitating a weighted approach to better assess and prioritize the risks associated with different vulnerabilities.

We further explored the smart contracts we extracted, starting with 5,762 unique contracts. Figure \ref{smart_contract_versions_histogram} shows that nearly 74\% have only one version, while only 0.28\% have more than 10 versions. The dominance of single-version contracts is likely due to deployment costs, but the reasons for multiple versions remain unclear and warrant further study.

\begin{figure}[htb]
    \centering
    \includegraphics[width=0.8\columnwidth]{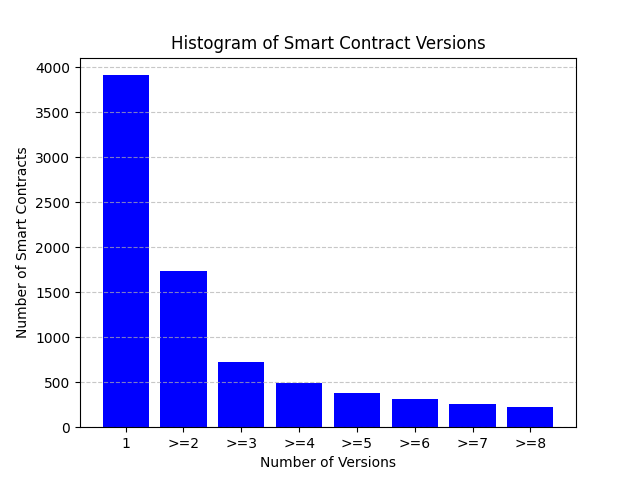}
    \caption{Versioning in smart contracts}
    \label{smart_contract_versions_histogram}
\end{figure}

\section{Conclusion and future works}
We developed Ethstractor, a tool to extract versioned smart contracts deployed on the Ethereum blockchain. This tool can help both academics and industry professionals analyze smart contract development and improve security auditing. We explored the relationships between different versions of Ethereum smart contracts, assessed the effectiveness of code metrics as vulnerability indicators, and examined the persistence of self-admitted technical debt. Our findings show that smart contract versions can be effectively traced through deployment metadata, but code metrics are ineffective as vulnerability predictors. Surprisingly, newer contract versions did not improve in mitigating vulnerabilities. Additionally, our analysis revealed that self-admitted technical debt is rare and seldom resolved, with only 6.49\% of such debts addressed in subsequent versions.
For future work, we aim to enhance our understanding of smart contract vulnerabilities by quantifying their severity and impact, and examining correlations with code metrics to identify which metrics can accurately predict specific vulnerability categories.

\bibliographystyle{IEEEtran}
\bibliography{IEEEabrv,refs}
\end{document}